\documentclass[prl,showpacs,twocolumn,superscriptaddress,floatfix,amsmath]{revtex4}
\usepackage[latin1]{inputenc}
\usepackage[english]{babel}
\usepackage{graphicx}
\usepackage{psfrag}
\usepackage{amssymb}
\usepackage{amsmath}
\usepackage{amscd}
\usepackage{eucal}
\usepackage{color}
\usepackage{bm}
\newcommand{\bi}{\bibitem}
\newcommand{\ignore}[1]{\relax}

\usepackage{epstopdf}
\DeclareMathAlphabet{\mathpzc}{OT1}{pzc}{m}{it}

\newcommand{\e}{{\rm e}}
\newcommand{\rmd}{{\rm d}}
\newcommand{\rmi}{{\rm i}}

\newcommand{\de}{\delta}

\newcommand{\veps}{\varepsilon}

\newcommand{\tD}{\tau_{\rm D}}
\newcommand{\NL}{N_{\rm L}}

\newcommand{\NR}{N_{\rm R}}
\newcommand{\NS}{N_{\rm S}}
\newcommand{\NT}{N_{\rm T}}

\definecolor{DarkGreen}{rgb}{0,0.7,0}

\begin{document}

\title{Controlling the Sign of Magnetoconductance in Andreev Quantum Dots}
\author{Robert S.~Whitney}
\affiliation{Institut Laue-Langevin,
6 rue Jules Horowitz, BP 156, 38042 Grenoble, France}
\author{Ph.~Jacquod}
\affiliation{Physics Department,
   University of Arizona, 1118 E. 4$^{\rm th}$ Street, Tucson, AZ 85721, USA}
\date{December 8, 2009}
\begin{abstract}
We construct a theory of coherent transport through a ballistic quantum dot coupled to a superconductor.
We show that the leading-order quantum correction to the two-terminal conductance of these
Andreev quantum dots may
change sign depending on (i) the number of channels carried by the normal leads or (ii) 
the magnetic flux threading the dot. 
In contrast, spin-orbit interaction may affect the magnitude of the correction, 
    but not always its sign.
Experimental signatures of the effect include a non-monotonic
magnetoconductance curve and a transition from an insulator-like to a metal-like temperature dependence of the conductance. 
Our results are applicable to ballistic or disordered dots.
\end{abstract}
\maketitle{}

{\bf Introduction.}
Low temperature experiments on diffusive metals coupled to superconductors
have reported large
interference effects analogous to coherent backscattering,
weak-localization and Aharonov-Bohm oscillations~\cite{Vanson87,Pet93,Cou96,Har97,Eom98+Jia05,Par03,Bau08},
one to two orders of magnitude 
above the universal amplitude ${\cal O}(e^2/h)$ they have in purely metallic mesoscopic
conductors~\cite{Akkermans}.
In some cases, a weak localization-like behavior, in the form
of positive magnetoconductance near zero field is observed~\cite{Par03,Bau08},
but often one sees negative magnetoconductance~\cite{Pet93,Cou96,Har97,Eom98+Jia05}.
Theoretical works predict that Andreev reflection from the superconductor
induces this large quantum correction to transport~\cite{Spi82,Bee95}.
The general expectation is that this correction is similar to a
magnified weak-localization correction, in that its sign is determined by the presence or 
absence of spin-orbit interaction (SOI)~\cite{Akkermans,Hik80}. 
In this paper we revisit this issue, and find that 
this interference correction has very different properties from weak-localization.  
In particular, we show that both the specific lead-geometry and an applied magnetic flux
can reverse its sign, while SOI need not.

\begin{figure}
\includegraphics[width=\columnwidth]{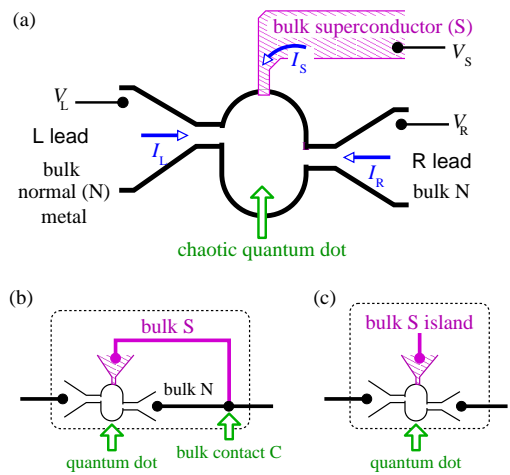}
\caption{\label{Fig:circuits}
(a) A two-dimensional Andreev quantum dot in a three-terminal geometry, with 
two normal (N) and one superconducting (S) lead. (b,c) The two possible two-terminal set-ups
obtained from such a dot. Either (b) the S lead is contacted to one of the N leads,
or (c) the S lead is floating.}
\end{figure}

\begin{figure}
\includegraphics[width=6.3cm]{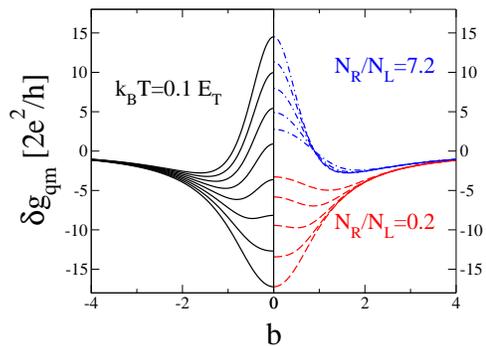}
\caption{\label{Fig:magnetoconductance}
Magnetoconductance curves for the set-up of Fig.~\ref{Fig:circuits}b. Left panel:
$k_{\rm B}T = 0.1 E_{\rm T}$, and
$N_{\rm R}/N_{\rm L}=n+0.2$, $n=0,1,2,...7$ (from bottom to top).
Right panel: $N_{\rm R}/N_{\rm L}=0.2$ (dashed red) and $7.2$ (dot-dashed blue), for
$k_{\rm B}T / E_{\rm T}=0.1$, 1, 2, 4 and 8 (dashed: from bottom to top; dot-dashed: from top to bottom).
For both panels, the vertical axis gives $\delta g_{\rm qm}$ in units of the conductance
quantum $2e^2/h$ with channel numbers chosen
such that $\NS^2\NL^2/(\NL+\NR)^3=5$ in all instances. 
Note the crossover from monotonic to non-monotonic behavior of the magnetoconductance
as $T$ increases, for $\NR/\NL=0.2$ (dashed red curves).}
\end{figure}

Andreev reflection~\cite{And64} is the dominant low energy process at the interface between a metal and a superconductor. It involves an electron (hole) being retroreflected as a hole (electron) and retracing the path previously followed by the electron (hole).
In this article, we extend the trajectory-based semiclassical theory to include Andreev reflection,
analyze the conductance of  
a two-dimensional ballistic quantum dot coupled to one superconducting lead and two normal leads, 
as in Fig.~\ref{Fig:circuits}. 
We dub this system an {\it Andreev quantum dot}.
We arrive at the surprising conclusion that the interference effects
can be reversed from localizing to antilocalizing
by changing the widths of the normal leads,
or by threading a fraction of a magnetic flux quantum through the dot.
In contrast SOI need not cause such a reversal.
This is very different from weak-localization in purely metallic conductors, whose sign is solely determined by the presence or absence of SOI~\cite{Akkermans,Hik80,caveat}. 
We predict two clear experimental signatures of these interference effects
in the form of non-monotonic magnetoconductance curves 
(see Fig.~\ref{Fig:magnetoconductance}) and 
a transition from
an insulator-like to a metal-like temperature-dependence of the conductance
as one changes the magnetic field or the ratio of the lead widths. This transition occurs because
thermal averaging destroys quantum interferences, thus depending on the sign of the
effect, the conductance increases or decreases by many times $e^2/h$ as the temperature
increases.

\begin{figure*}[t]
\includegraphics[width=0.96\textwidth]{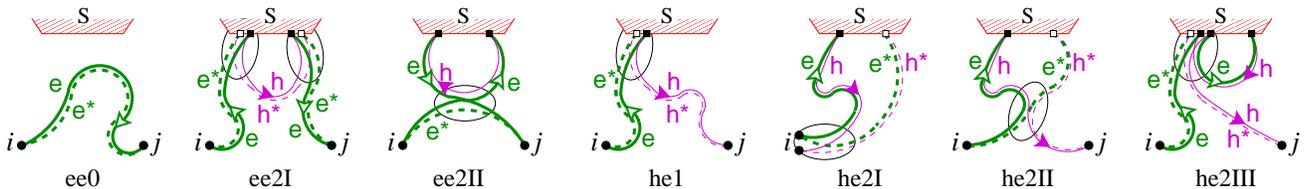}
\caption{\label{fig:T-all} 
Contributions to $\langle T_{ij}^{\rm ee}\rangle$ (first three)
and $\langle T_{ij}^{\rm he} \rangle$ (last four) considered in this letter.
Thick green (thin violet) paths indicate electrons (holes),
dashed lines indicate complex-conjugated amplitudes.
Normal leads are labelled $i,j$ while the S lead is superconducting.
The contributions are classified by the number of uncorrelated Andreev reflections
(ee0 has none, ee2I and ee2II both have two).
The full (open) squares on the S lead indicate a factor of $\eta$ ($\eta^*$)
and the ellipses mark encounters.}
\end{figure*}

{\bf Semiclassical transport with superconductivity.}
According to the scattering approach to transport,
the current in normal lead $i$ is given by~\cite{Lam93}
\begin{eqnarray}\label{eq:current}
I_i &=&{2e \over h} 
\int_0^\infty \rmd \varepsilon 
\sum_j \left[2 N_i\delta_{ij}-T_{ij}^{\rm ee}+T_{ij}^{\rm he} 
-T_{ij}^{\rm hh}+T_{ij}^{\rm eh}
\right] \nonumber \\
&& \hspace{2.2cm} \times (-\partial f/\partial \varepsilon)(\mu_j-\mu_{\rm S}) \, ,
\end{eqnarray}
where $\mu_j$ is the chemical potential of normal (N) lead  $j$ 
and $\mu_{\rm S}$ of all the superconducting (S) leads.
The Fermi-Dirac distribution,
$f(\varepsilon)$, has $\varepsilon$ measured from $\mu_{\rm S}$.
Here we use trajectory-based semiclassics 
to find the scattering probability $T_{ij}^{\alpha \beta}$  to go from quasiparticle $\beta={\rm e,h}$ (electron, hole) in  lead $j$ to quasiparticle
$\alpha$ in lead $i$.
Extending trajectory-based semiclassics \cite{Ric02,Whi06,Essen-papers,Bro06,Ber08} to 
include Andreev reflection,
one has~\cite{Goo08}
\begin{eqnarray}\label{semicl-tr}
T_{ij}^{\alpha\beta} &=& \frac{1}{2 \pi \hbar} \int_j \rmd y_0 \int_i \rmd y
\sum_{\gamma1,\gamma2} A_{\gamma1} A_{\gamma2}^* \exp[i \delta S/\hbar]
\, .\qquad
\end{eqnarray}
This expression sums over all  classical 
trajectories $\gamma1$ and $\gamma2$ entering the cavity 
at $y_0$ on a cross-section
of lead $j$ and exiting at $y$ on a cross-section of lead $i$, while
converting a $\beta$ quasiparticle into an $\alpha$  quasiparticle. 
The phase $\delta S = S_{\gamma1}-S_{\gamma2}$
gives the difference in action phase accumulated along $\gamma1$ and $\gamma2$,
and $A_\gamma$ gives the stability of the trajectory $\gamma$.
In contrast to Ref.~\cite{Goo08}, we consider the physically more prevalent situation of 
an Ehrenfest time negligible against the dwell time $\tau_{\rm D}$ inside the dot. In that case, 
even with perfect Andreev reflection, 
quantum uncertainties combined with the chaotic dynamics
make the retroreflected quasiparticle 
diverge from the incoming quasiparticle path well before
it leaves the dot~\cite{Lar68}. Therefore classical paths
undergoing Andreev reflections consist of electron and hole
segments that do not necessarily retrace each other all the way.
For transmission probabilities $\langle T_{ij}^{\alpha \beta} \rangle$ 
averaged over energy or dot shape,
we must pair the paths $\gamma1$ and $\gamma2$ in Eq.~(\ref{semicl-tr}) 
in ways that render their action phase 
difference stationary.
To do this we either pair a path with a complex conjugate 
path, e-e$^*$ or h-h$^*$, or we pair an electron path with a hole 
path, e-h or e$^*$-h$^*$.
Path-pairs can meet and swap pairings at 
{\it encounters}, as shown in Fig.~\ref{fig:T-all}. Following Ref.~\cite{Whi06}
we distinguish between encounters
that lie entirely inside the dot and those that touch a lead.

{\bf Feynman rules.}
Contributions relevant to current noise 
in purely metallic samples~\cite{Whi06,Essen-papers,Bro06,Ber08},
become relevant for the calculation of the current itself in the presence of S leads
when they can be made from
only two classical trajectories with some segments as electron and others as holes.
From Refs.~\cite{Whi06,Bro06,Ber08} and the above considerations, we derive
the following Feynman rules for calculating transmission
through an Andreev quantum dot.  
The  dot is connected to normal and superconducting leads, 
each carrying $N_i\gg1$ and $N_{{\rm S}j}\gg 1$ transport channels respectively, 
and we write $N_{\rm T}= \sum_i N_i+ \sum_j N_{{\rm S}j}$ 
For a perpendicular magnetic field, $b=B/B_{\rm c}$, measured in units of the field 
$B_{\rm c} \simeq (h/e A)(\tau_0/\tau_{\rm D})^{1/2}$ 
that breaks time-reversal (TR) symmetry in a quantum dot of area $A$ 
with time of flight  $\tau_0$, the Feynman rules read:

\noindent (i) An e-e$^*$ or h-h$^*$ path-pair gives a factor of 
$[N_{\rm T}(1+\chi b^2)]^{-1}$, with $\chi=1$ for time-reversed paths
and $\chi=0$ otherwise.\\
(ii) An e-h or  e$^*$-h$^*$ path-pair gives
$N_{\rm T}^{-1}\times (1\pm\rmi 2\varepsilon\tau_{\rm D}+\chi b^2)^{-1}$,
with upper (lower) sign for e-h (e$^*$-h$^*$).\\
(iii) An encounter inside the dot and connecting e,e$^*$, h and h$^*$ paths
(as in he2II) gives a factor $-N_{\rm T}$.\\
(iv) An encounter inside the dot and connecting e, e, e$^*$ and h paths (as in ee2II) 
gives a factor of  $-N_{\rm T}(1+ \rmi2 \varepsilon\tD+b^2)$; this factor is complex conjugated (c.c.)
for an encounter connecting e, e$^*$, e$^*$ and h$^*$ paths.\\
(v) An encounter touching a  N lead $i$ (S lead $j$) 
gives a factor of $N_i$ ($N_{{\rm S}j}$).\\
(vi) A path-pair that ends at a N lead $i$ (S lead $j$), while not in an encounter,
gives a factor of $N_i$ ($N_{{\rm S}j}$).\\
(vii) Andreev reflections at S leads
involving e$\to$h give a factor of $\eta\, \e^{- \rm i \Phi_{{\rm S}j}}$ while those involving h$\to$e give a factor of $\eta\, \e^{\rm i \Phi_{{\rm S}j}}$
(e$^*\to$h$^*$ and  h$^*\to$e$^*$ give the c.c.~of these factors),
where $\Phi_{{\rm S}j}$ is the S phase on lead $j$, and 
$\eta=\exp[- {\rm i arcos} \,(\varepsilon/\Delta)]$ is
the Andreev reflection phase.

\noindent We note that these rules
equally follow from random-matrix theory \cite{Essen-papers}.

In our analysis of the consequences of these rules,
we consider temperatures well below the
superconducting gap $\Delta$ where $\eta=-\rmi$, 
and consider a single S lead (setting 
$\Phi_{\rm S}=0$ without loss of generality). 
The rules indicate that a path-pair going from encounter
to encounter reduces the contribution by a factor of ${\cal O}[N_{\rm T}]$.
Thus to leading order in $N_{\rm T}$, we can neglect 
such (weak localization) contributions.
This does not restrict the number of encounters, because
the price to add an encounter 
whose additional legs go to S leads
is  ${\cal O}[(N_{\rm S}/N_{\rm T})^2]$. We therefore 
take $\NS/\NT \ll 1$, and 
expand in the number of uncorrelated Andreev reflections.

Restricting ourselves to
${\cal O}[(N_{\rm S}/N_{\rm T})^2]$, we need to 
consider the contributions shown in Fig.~\ref{fig:T-all}
involving no more than two uncorrelated Andreev reflections.
The contributions to $\langle T^{\rm ee}_{ij}\rangle$ are
\begin{subequations}\label{eq:contribs-ee}
\begin{eqnarray}
\langle T^{\rm ee0}_{ij}\rangle &=& N_i N_j/N_{\rm T},  
\\
\langle T^{\rm ee2I}_{ij} \rangle
&=& N_i N_j \NS^2/N_{\rm T}^3,
\\
\langle T^{\rm ee2II}_{ij} \rangle
&=& 
{2N_i N_j \NS^2\over N_{\rm T}^3} \,{\rm Re} \left[(1+b^2 +\rmi 2\veps \tD^2)^{-1}\right].
\end{eqnarray}
\end{subequations}
The contributions to $\langle T^{\rm he}_{ij}\rangle$ are 
\begin{subequations}\label{eq:contribs-he}
\begin{eqnarray}
\langle T^{\rm he1}_{ij} \rangle
&=& N_i N_j \NS/N_{\rm T}^2, 
\\
\langle T^{\rm he2I}_{ij}  \rangle
&=& \de_{ij}\,  N_i \NS^2/\left[ N_{\rm T}^2
\big( (1+b^2)^2+4 \veps^2 \tau^2_{\rm D}\big) \right],
\\
\langle T^{\rm he2II}_{ij}  \rangle
&=& -N_i N_j \NS^2/\left[N_{\rm T}^3 \big((1+b^2)^2+4 \veps^2 \tau^2_{\rm D}\big) \right], \qquad
\ \\
\langle T^{\rm he2III}_{ij} \rangle
&=& -\langle T^{\rm ee2II}_{ij}\rangle.
\end{eqnarray}
\end{subequations}
Semiclassics gives
$\langle T_{\rm ij}^{\rm hh}\rangle= \langle T_{\rm ij}^{\rm ee}\rangle$ 
and  $\langle T_{\rm ij}^{\rm eh}\rangle= \langle T_{\rm ij}^{\rm he}\rangle$.
These contributions preserve unitarity up to and including
${\cal O}[(N_{\rm S}/\sum_i N_i)^2]$.

{\bf Set-up with an S lead.}
We first consider the situation where the S lead's potential 
is fixed externally.
This may be the
three-terminal device of Fig.~\ref{Fig:circuits}a with 
both the R and S leads grounded, while the
L lead is biased at electrochemical potential $\mu_{\rm L}=eV$.
Alternatively this may be the two-terminal device  in Fig.~\ref{Fig:circuits}b 
with the S and R leads join at a bulk
contact (with contact conductance vastly greater than the dot),
a macroscopic distance away from the dot. 
In either case, 
the L lead current is
$I_{\rm L}  =  (2 e^2/h) \left[g_{\rm cl} 
+ \delta g_{\rm qm}(T,b)\right] V$, where we define
a dimensionless classical conductance 
$g_{\rm cl}= N_{\rm L}(N_{\rm R}+2N_{\rm S})
/(N_{\rm L}+N_{\rm R}+2N_{\rm S})$~\cite{footnote:drude}.
For $N_S \ll N$ the quantum interference correction is
\begin{equation}\label{eq:g_q-correction}
\delta g_{\rm qm} =
\frac{N_{\rm L}[N_{\rm R}- 4 N_{\rm L}(1+b^2)] N_{\rm S}^2}
{(N_{\rm L}+N_{\rm R})^3} f(T,b)  + \de g_{\rm wl}.
\end{equation}
In the regime of experimental interest the weak-localization correction in the absence of 
the S lead $\de g_{\rm wl} \simeq  -N_{\rm L}N_{\rm R}/[(N_{\rm L}+N_{\rm R})^2(1+b^2)]$ is small enough to neglect  
(as in Fig.~\ref{Fig:magnetoconductance}).
The  $\varepsilon$-integral in Eq.~(\ref{eq:current}) 
with  $\langle T_{ij}^{\alpha \beta} \rangle$ in Eqs.~(\ref{eq:contribs-ee},\ref{eq:contribs-he}) leads to 
$f(T,b) = \alpha \, \zeta(2,1/2+(1+b^2) \alpha)$, with
$\alpha=E_{\rm T}/4 \pi k_{\rm B} T$ for
a Thouless energy $E_{\rm Th} = \hbar/\tau_{\rm D}$, 
and the 
generalized $\zeta$-function $\zeta(2,x)=\int_0^\infty t \exp[-x t]/(1-\exp[-t]) {\rm d}t$.
This gives the two asymptotics $f(T\rightarrow \infty,b) \rightarrow \pi E_{\rm T}/[8 k_{\rm B}T(1+b^2)]$
and $f(T\rightarrow 0,b) \rightarrow 1/(1+b^2)^2$.

At zero temperature, we find three regimes
for $\de g_{\rm qm}$:

\noindent (a) For $\NR < 2 \NL $, $\de g_{\rm qm} < 0$ for all values of $b$,
and gives a monotonic magnetoconductance curve. 

\noindent (b) For $2\NL < \NR < 4\NL$, $\de g_{\rm qm}<0$ for all $b$,
but gives a non-monotonic magnetoconductance, with 
a minimum at $b^2=  (\NR-2\NL)/(2\NL)$.

\noindent (c) For $\NR > 4\NL$, $\de g_{\rm qm}>0$ at small $b$, but becomes negative
for $b^2> (\NR-4\NL)/(4\NL)$, and then goes to zero for large $b$.
As in (b), the curve is non-monotonic with minima at $b^2=  (\NR-2\NL)/(2\NL)$.

\noindent These different regimes persist at finite temperature as is illustrated in
Fig.~\ref{Fig:magnetoconductance}, however,
the boundary between (a) and (b),
as well as the positions of the minima of the magnetoconductance curves
are $T$-dependent.
The conductance exhibits
a metal-like (insulating-like) behavior in the form of a decrease (increase) of the conductance
with $T$, depending on the sign of $(\NR -4(1+b)^2\NL)$.
This sign can easily be changed, whenever one has control over the lead widths 
or the magnetic flux. Remarkably, a monotonic magnetoconductance may 
become non-monotonic upon increase of the temperature (dashed red curves in Fig.~\ref{Fig:magnetoconductance}).

{\bf Set-up with an S island.}
In the second of the two possible two-terminal set-ups, Fig.~\ref{Fig:circuits}c),
the quantum dot is connected to
a superconducting island, whose chemical potential is floating, and adapts itself to a value guaranteeing
current conservation, $I_{\rm L}=-I_{\rm R}$. 
Using the  expression in Ref.~\cite{Lam93} for the two-terminal 
conductance in terms of the transmission probabilities, $T^{\alpha\beta}_{ij}$,
we obtain $g= g^{\rm isl}_{\rm cl} + \delta g^{\rm isl}_{\rm qm}(T,b)$
where $g^{\rm isl}_{\rm cl}= N_{\rm L}N_{\rm R}/(N_{\rm L}+N_{\rm R})$
and $\delta g^{\rm isl}_{\rm qm} =
N_{\rm L}N_{\rm R} N_{\rm S}^2(N_{\rm L}+N_{\rm R})^{-3}\,f(T,b)$.
This reproduces the random matrix theory result \cite{Bee95} to leading order in
$[N_{\rm S}/(N_{\rm L}+N_{\rm R})]^2$.
This quantum correction always increase the conductance (antilocalization)
by a parametrically large amount (many $e^2/h$), with a monotonic magnetoconductance curve.

{\bf Mesoscopic conductance fluctuations and current noise.}
Ref.~\cite{Bro96} used random matrix theory to show that conductance fluctuations remain
${\cal O}(e^2/h)$ in the presence of superconductivity.
Our Feynman rules reproduce this result.
Contributions to ${\rm var}[g]$ are the product of any two contributions in Fig.~\ref{fig:T-all} connected by encounters. 
Since path-pairs are not swapped at entrance and exit, the connection must involve at least two additional encounters with four additional path-pairs, and the resulting contribution 
behaves as $N_{\rm T}^{-2}$ times the average conductance squared. 
This is at most ${\cal O}[N_{\rm T}^0]$,
thus the quantum corrections to the 
average conductance are parametrically larger than the conductance fluctuations, 
and are therefore easily observable.

The S contact also leads to e-h contributions to the current-noise~\cite{Ana96}. The Feynman rules show that they are  ${\cal O}[N_{\rm T}(N_{\rm S}/N_{\rm T})^n]$
for $n \ge 1$ and are thus smaller than the ${\cal O}[N_{\rm T}]$ e-e contributions which give the noise in the absence of an S lead. Therefore, to leading order
in $N_{\rm S}/N_{\rm T}$, the parametric magnitude of the zero-frequency current-noise
is unaltered by the presence of the S lead.

{\bf Effect of SOI.}
Spin-orbit interaction (SOI) can be treated as rotating the spin along otherwise unchanged classical trajectories, 
multiplying Eq.~(\ref{semicl-tr}) by ${\rm Tr} [d_{\gamma1} d_{\gamma2}^\dagger]$,
where $d_{\gamma i}$ is the SU(2)-phase of path $\gamma i$~\cite{Mat92}. 
For ee0, ee2I and he1,
this gives a factor of two for spin-degeneracy, 
because $d_{\gamma1} = d_{\gamma2}$.
However for ee2II and he2III
it gives ${\rm Tr} [d_1^2]$, 
and for he2I and he2II it gives ${\rm Tr} [d_1^2 d_2^2]$,
where $d_1,d_2$ are statistically independent random SU(2) phases.
When the SOI time is shorter than $\tau_{\rm D}$, one averages 
uniformly over the SU(2) phases~\cite{Zai05}, which
multiplies ee2II and he2III by $-1/2$, and he2I and he2II by $1/4$.
Taking $T=0$ and neglecting $\de g_{\rm wl}$ for simplicity, we find that
\begin{subequations}
\begin{eqnarray}
\delta g_{\rm qm} &=&    {\NL \, [(1-2/\beta)^2\NR + 4(1-2/\beta)\NL] \, \NS^2 \over (\NL+\NR)^3}, \qquad
\\
\delta g^{\rm isl}_{\rm qm} 
&=& {(1-2/\beta)^2 N_{\rm L}N_{\rm R} N_{\rm S}^2/ (N_{\rm L}+N_{\rm R})^3},
\end{eqnarray}
\end{subequations}
for the three standard symmetry classes,
$\beta=1$ (TR symmetry without SOI),
 2 (no TR symmetry) and 4 (TR symmetry with SOI).
Note the presence of the same symmetry prefactor $(1-2/\beta)$ as for weak localization
without superconductivity.
Thus with SOI ($\beta=4$), both $\delta g_{\rm qm}$ and $\delta g^{\rm isl}_{\rm qm}$ always enhance conductance.
Therefore, SOI must be absent for a sign change of $\delta g_{\rm qm}$ with lead width.  Turning on SOI (going from $\beta=1$ to $\beta=4$)
never changes the sign of $\delta g^{\rm isl}_{\rm qm}$ but changes
the sign of $\delta g_{\rm qm}$  for $\NR < 4\NL$. 

{\bf Concluding remarks.}
The derivation outlined here is for ballistic quantum dots, 
however the Feynman rules that we analyze apply to any system well-modelled by random matrix theory.  Thus our results are equally applicable to 
disordered dots.
We also expect
qualitatively similar behaviors in diffusive metals coupled to superconductors at intermediate
temperatures, $k_{\rm B} T \sim E_{\rm T}$. Work in this regime is in progress.

Upon completion of this work, we noted Ref.~\cite{Regensburg-gang} which uses a 
somewhat similar methodology as ours in closed Andreev billiards.
RW thanks L.~Saminadayar and C.~B\"auerle for stimulating 
discussions, and access to their data~\cite{Bau08}. PJ thanks the
Physics Department of the Universities of Geneva and Basel
as well as the Aspen Center for Physics for their hospitality at various stages of
this project and  
acknowledges the support of the National Science Foundation under Grant
No. DMR-0706319.



\begin{thebibliography}{99}
\bibitem{Vanson87} 
P.C. van Son, H. van Kempen, and P. Wyder, Phys. 
Rev. Lett. {\bf 59}, 2226 (1987);  J. Phys. 
F {\bf 18}, 2211 (1988). 

\bi{Par03} A. Parsons, I.A. Sosnin, and V.T. Petrashov, Phys. Rev. B {\bf 67}, 140502(R) (2003).

\bi{Pet93} 
V.T. Petrashov, V.N. Antonov, P. Delsing, and R. Claeson, Phys. Rev. Lett. {\bf 70}, 347 (1993).

\bi{Cou96}
H. Courtois, Ph. Gandit, D. Mailly, and B. Pannetier,
Phys. Rev. Lett. 76, 130 (1996).

\bi{Har97}
S.G. den Hartog, B.J. van Wees, Yu.V. Nazarov,
T.M. Klapwijk, and G. Borghs, Phys. Rev. Lett. {\bf 79}, 3250 (1997).

\bi{Eom98+Jia05}
J. Eom, C.-J. Chien, and V. Chandrasekhar, Phys. Rev. Lett. {\bf 81}, 437 (1998);
Z. Jiang and V. Chandrasekhar, Phys. Rev. B {\bf 72}, 020502(R) (2005).

\bi{Bau08}  C.~B\"auerle and L.~Saminadayar, Private Comm. (2008).

\bi{Akkermans} E. Akkermans and G. Montambaux, {\it Mesoscopic Physics of
Electrons and Photons} (Cambridge University, Cambridge, 2007).

\bibitem{Spi82} B.Z.~Spivak and D.E.~Khmelnitskii, JETP Letters, {\bf 35}, 412 (1982).

\bibitem{Bee95}  C.W.J. Beenakker, J.A. Melsen, and P.W. Brouwer, Phys. Rev. B {\bf 51}, 13883 (1995).

\bi{Hik80} S. Hikami, A.I. Larkin and Y. Nagaoka, Prog. Theor. Phys. {\bf 63}, 707 (1980).

\bi{And64} A.F. Andreev, Sov. Phys. JETP 19, 1228 (1964).


\bi{caveat} We exclude multiterminal measurements where 
the resistance is not necessarily 
an extremum at zero field. See: M. B\"uttiker, Phys. Rev. Lett. {\bf 57}, 1761 (1986).


\bi{Lam93} C.~J. Lambert, J. Phys.: Cond. Mat. {\bf 5}, 707 (1993).


\bibitem{Ric02} K. Richter and M. Sieber, Phys. Rev. Lett. {\bf 89}, 
206801 (2002).


\bibitem{Whi06} 
R.S.~Whitney and Ph.~Jacquod, Phys.~Rev.~Lett.~{\bf 96}, 206804 (2006).

\bibitem{Essen-papers}
P.~Braun, S.~Heusler, S.~M\"uller,  and F.~Haake, 
J. Phys. A: Math. Gen. {\bf 39}, L159 (2006).

\bibitem{Bro06}
P.W.~Brouwer, and S.~Rahav,
Phys.~Rev.~B {\bf 74}, 085313 (2006). 


\bibitem{Ber08}
G.~Berkolaiko, J.M.~Harrison and M.~Novaes, 
J.~Phys.~A {\bf 41}, 365102 (2008). 


\bibitem{Goo08} M.C. Goorden, Ph. Jacquod, and J. Weiss,
Phys. Rev. Lett. {\bf 100}, 067001 (2008); 
Nanotechnology {\bf 19}, 135401 (2008).

\bibitem{Lar68} A.I. Larkin and Yu.N. Ovchinnikov, Zh. Eksp. Teor. Fiz. {\bf 55},
2262  (1968)  [Sov. Phys. JETP {\bf 28}, 1200 (1969)].

\bibitem{footnote:drude}
This expression for $g_{\rm cl}$ is correct to all orders in $N_{\rm S}/N_{\rm T}$.

\bibitem{Bro96} P.W. Brouwer and C.W.J. Beenakker, Phys. Rev. B 
{\bf 54}, R12705 (1996).

\bibitem{Ana96} M.P. Anantram and S. Datta,  Phys. Rev. B {\bf 53}, 16390 (1996).

\bibitem{Mat92} H. Mathur and A.D. Stone, Phys. Rev. Lett. {\bf 68}, 2964 (1992).

\bibitem{Zai05} O. Zaitsev, D. Frustaglia, and K. Richter, Phys. Rev. B {\bf 72}, 155325 (2005).

\bibitem{Regensburg-gang}
J.~Kuipers, C.~Petitjean, D.~Waltner, and K.~Richter,
preprint --  arXiv:0907.2660.
\end{thebibliography}
\end{document}